\documentclass[11pt]{jpp}
\usepackage{epstopdf, epsfig}

\newcommand{\be}{\begin{displaymath}}
\newcommand{\bn}{\begin{equation}}
\newcommand{\bea}{\begin{eqnarray*}}
\newcommand{\eea}{\end{eqnarray*}}
\newcommand{\en}{\end{equation}}
\newcommand{\ee}{\end{displaymath}}

\renewcommand{\p}{\partial}
\newcommand{\lang}{\left\langle}
\newcommand{\rang}{\right\rangle}

\newcommand{\simlt}{\:{\raisebox{-1.5mm}{$\stackrel
{\textstyle{<}}{\sim}$}}\:}

\shorttitle{Available energy}
\shortauthor{P. Helander}

\title{Available energy of magnetically confined plasmas}

\author{Per Helander\aff{1}
  \corresp{\email{per.helander@ipp.mpg.de}}
}

\affiliation{\aff{1}Max-Planck-Institut f\"ur Plasmaphysik, D-17491 Greifswald, Germany
}

\begin{document}

\maketitle

\begin{abstract}
The concept of available energy of a collisionless plasma is discussed in the context of magnetic confinement. The available energy quantifies how much of the plasma energy can be converted into fluctuations (including nonlinear ones) and is thus a measure of plasma stability, which can be used to derive linear and nonlinear stability criteria without solving an eigenvalue problem. In a magnetically confined plasma, the available energy is determined by the density and temperature profiles as well as the magnetic geometry. It also depends on what constraints limit the possible forms of plasma motion, such as the conservation of adiabatic invariants and the requirement that the transport be ambipolar. A general method based on Lagrange multipliers is devised to incorporate such constraints in the calculation of the available energy, and several particular cases are discussed. In particular, it is shown that it is impossible to confine a plasma in a Maxwellian ground state relative to perturbations with frequencies exceeding the ion bounce frequency.

\end{abstract}

\section{Introduction}

In a previous publication [\cite{Helander-2017-b}], hereafter referred to as I, a quantity called the `available energy' was proposed as a measure of nonlinear plasma stability. The available energy is defined as the difference between the total plasma energy and the lowest value it could possibly attain under the various constraints that limit the motion of the plasma. It thus measures how much energy can, in principle, be released from the plasma in the form of linear and nonlinear instabilities. It is analogous to a quantity called the 'available potential energy' in meteorology, which is defined as the difference between the potential energy of the atmosphere and the minimum attainable by any adiabatic redistribution of mass [\cite{Lorenz-1955}]. This analogy is made mathematically explicit in the Appendix. 

In the present paper, the concept of available energy is explored further. Its simplest version was considered many years ago by \cite{Gardner-1963}, who identified the lowest-energy state of a Vlasov plasma under the sole constraint that its motion should satisfy Liouville's theorem. As pointed in I, the concept of available energy becomes much more interesting and relevant if one also accounts for the fact that adiabatic invariants are usually conserved. 

Indeed, most important types of instabilities and turbulence affecting magnetically confined plasmas are characterised by frequencies below the ion gyrofrequency $\Omega_i$, implying that the magnetic moment,
	$$\mu_a = \frac{m_a v_\perp^2}{2B}, $$
of each particle is conserved. Here $m_a$ denotes the mass of the species $a$ in question and $v_\perp$ the speed perpendicular to the magnetic field ${\bf B} = B {\bf b}$. If, moreover, the frequency is below the electron bounce/transit frequency $\omega_{be}$, the parallel adiabatic invariant
	$$ J_a = \int m_a v_\| dl $$
is conserved for the electrons ($a=e$), where $l$ denotes the arc length along the magnetic field and 
	$$ m_av_\| = \sqrt{2 m_a (H_a - \mu_a B - e_a \Phi)}, $$
the parallel momentum. The energy 
	$$ H_a = \frac{m_a v^2}{2} + e_a \Phi $$
(where $\Phi$ denotes the electrostatic potential) and $\mu_a$ are kept constant in the integration, which is taken between two consecutive bounce points, defined by $v_\| = 0$.
 
For drift-wave instabilities and turbulence (e.g., ion-temperature-gradient and trapped-electron-modes), the characteristic frequency $\omega$ usually satisfies the ordering
	\bn \omega_{bi} \sim \omega \ll \omega_{be} \simlt \Omega_i \ll \Omega_e, 
	\label{ordering}
	\en
so that $\mu_e$, $\mu_i$ and $J_e$ are conserved, but not $J_i$. (The subscript $i$ refers to the ions, which will be assumed to be singly charged.) Different constraints thus pertain to the electron and ion distribution functions, and their lowest-energy states will therefore be different. However, the plasma must remain quasineutral, which imposes an additional constraint on the evolution of the plasma that was not accounted for in I.

The aim of the present paper is to discuss how this and other relevant constraints affect the available energy of magnetically confined plasmas. The energy available to fluctuations subject to constraints such as Eq.~(\ref{ordering}) is lower than that available to less constrained fluctuations, and our goal is to develop a general method for incorporating various types of constraints. We begin in Section 2 by briefly recapitulating the basic results of I, deriving them in a particularly expeditious way. This derivation naturally lends itself to the inclusion of additional constraints, and in the following several sections we show how these can be incorporated. In Section 3, the available energy of a single plasma species with conserved $\mu$ and $J$ is calculated; in Sections 4 and 5 the constraint of a fixed density profile is explored; and in Section 6 ground states of a plasma satisfying the ordering (\ref{ordering}) are investigated, where different constrains apply to the ions and electrons. 

\section{Minimum-energy states}

As in I, let us first consider the distribution function $f({\bf x},t)$ of a single particle species governed by the Vlasov equation or any other kinetic equation satisfying the Liouville theorem,	
	$$ \frac{\p f}{\p t} + \dot x_k \frac{\p f}{\p x_k} = 0, $$
	$$ \frac{\p}{\p x_k} (\dot x_k \sqrt{g}) = 0, $$
where the phase-space coordinates ${\bf x}$ are arbitrary except that the Jacobian $\sqrt{g}$ is assumed to be independent of time. Summation over repeated indices is understood. Liouville's theorem implies that the flux of particles in phase space is incompressible, so that, for any positive number $\phi$, the volume of the set of points satisfying $f({\bf x},t) > \phi$ is constant, i.e.,
	\bn \frac{\p}{\p t} H[f({\bf x},t),\phi] = 0, 	
	\label{H0}
	\en
where the functional $H$ is defined by
	$$ H[f({\bf x},t),\phi] = \int \Theta[f({\bf x},t) - \phi] \sqrt{g} \; d{\bf x} $$
and $\Theta$ denotes the Heaviside step function.

Now let $\epsilon({\bf x})$ denote the particle energy as a function of the phase-space coordinates $\bf x$. (Typically $\epsilon = mv^2/2$, where $v$ denotes speed.) The total energy associated with the distribution function $f$ is then
	$$ E[f] = \int \epsilon f \sqrt{g} \; d{\bf x}, $$
and we may ask for the minimum energy that can be attained starting from some initial state $f({\bf x},0)$. We denote the energy-minimising distribution function by $f_0({\bf x})$ and note that, because of the constraint (\ref{H0}), we require
	\bn H[f_0({\bf x}),\phi] = H[f({\bf x},0),\phi] \equiv H_0(\phi). 
	\label{H}
	\en
We can account for this constraint by introducing a continuous set of Lagrange multipliers $\lambda(\phi)$ and minimising the functional
	\bn W[f_0,\lambda] = E[f_0] + \int_0^\infty \lambda(\phi) d\phi
	\left[ \int \Theta[f_0({\bf x}) - \phi] \sqrt{g} \; d{\bf x} - H_0(\phi) \right]. 
	\label{W}
	\en
The minimisation with respect to $\lambda$ then ensures that the constraint (\ref{H}) is satisfied, whereas the variation with respect to $f_0$ gives
	$$ \delta W = \int \delta f_0 \left[ \epsilon + \lambda(f_0)\right] \sqrt{g} \; d{\bf x} = 0, $$
implying that $\lambda(f_0) = - \epsilon({\bf x})$. Inverting this relation, we come to the conclusion that the lowest-energy state can only depend on the phase-space coordinates through $\epsilon({\bf x})$, i.e., it must be possible to write $f_0$ as
	\bn f_0({\bf x}) = F[\epsilon({\bf x})] 
	\label{ground state}
	\en
for some function $F$. 

As originally shown by \cite{Gardner-1963}, this function should be monotonically decreasing. Otherwise, the distribution function does not correspond to a minimum-energy state, because the energy can be lowered by interchanging two neighbouring phase-space elements of equal volume $d\tau = \sqrt{g} d{\bf x}$ but different energies. To see this, let us denote the distribution function before the interchange by $f_0$ and afterwards by $f_1$. Then
	$$ f_1({\bf x}_b) = f_0({\bf x}_a), $$
	$$ f_1({\bf x}_a) = f_0({\bf x}_b), $$
where points in the two respective phase-space elements are denoted by ${\bf x}_{a,b}$. As a result of the interchange, the energy changes by the amount
	$$ \Delta E = \int \epsilon (f_1-f_0) \sqrt{g} d{\bf x}
	= -[\epsilon({\bf x}_a) - \epsilon({\bf x}_b)] [f_0({\bf x}_a) - f_0({\bf x}_b)] d\tau, $$
and becomes equal to 
	$$ \Delta E = -[\epsilon({\bf x}_a) - \epsilon({\bf x}_b)]^2 F_0'(\epsilon) $$
if the distribution function is of the form (\ref{ground state}). Hence it is clear that $F'(\epsilon)$ must be negative (or zero) everywhere in a lowest-energy state, which we called the {\em ground state} in I.  

Since $F$ is a monotonically decreasing function we have
	$$ \Theta\left[F(\epsilon({\bf x})) - \phi \right] = \Theta\left[w - \epsilon({\bf x}) \right], $$
where $F(w) = \phi$. It follows that this function is determined by the condition
	\bn H_0[F(w)] = \Omega(w), 
	\label{Ground-state eq}
	\en
for each $w \in [0,\infty)$, where
	$$ \Omega(w) = \int \Theta \left[w - \epsilon({\bf x}) \right] \sqrt{g} \; d{\bf x}, $$
measures the volume of the subset of phase space in which $\epsilon({\bf x}) < w$. Equation (\ref{Ground-state eq}) states that this volume must equal the volume in which the distribution function exceeds $F(y)$, as follows from Liouville's theorem. 

Equation (\ref{Ground-state eq}) determines the ground states of a plasma whose evolution is only constrained by Liouville's theorem. As demonstrated in I and discussed further below, it is not difficult to account for more constraints by using additional Lagrange multipliers. 

The available energy is defined as the difference in energy between the initial state and the ground state,	
	$$ A = \int (f - f_0) \epsilon \sqrt{g} \; d{\bf x}. $$
It evidently constitutes an upper bound on the energy that can be extracted from the plasma by linear and nonlinear instabilities, but this upper bound is in general unattainable. For instance, it does not account for the fact that the fluctuations created by the instability in question must also be contained in $f$. Moreover, because coherent waves tend to diffuse particles or because internal wave turbulence interacts incoherently with particles, particles tend to diffuse in phase space (looking at realistic granularity) rather than to conserve the phase space density.  Gardner restacking therefore overestimates the available free energy [\cite{Fisch-1993}, \cite{Hay-2015}].

\section{Conservation of $\mu$ and $J$}

As already remarked in the Introduction and discussed at length in I, constraints implied by the adiabatic invariants $\mu$ and $J$ can place severe limits on the available energy. We are particularly interested in the ordering (\ref{ordering}), according to which both these invariants are conserved for trapped electrons.\footnote{Circulating particles play no role in the section. In the ordering (\ref{ordering}) they will not exchange much energy with the fluctuations, as discussed further in Section 4.} In the present section, we calculate the ground state of any such particle species for which $\mu$ and $J$ are conserved. 

\subsection{The role of omnigenity}

For a plasma with magnetic surfaces, it is useful to write the magnetic field as ${\bf B} = \nabla \psi \times \nabla \alpha$, where $\psi$ labels the magnetic surfaces and $\alpha \in [0,2\pi)$ the different field lines thereon. The distribution function is then usefully considered as a function of the phase-space coordinates $(\psi, \alpha, \mu, J)$, and we shall take it to be  initially Maxwellian,
	\bn f(\psi, \alpha, \mu, J) = f_M \left[ \epsilon(\psi, \alpha, \mu, J), \psi \right]
	= n(\psi) \left(\frac{m}{2 \pi T(\psi)} \right)^{3/2} e^{-\epsilon(\psi, \alpha, \mu, J)/T(\psi)}, 
	\label{Maxwellian flux function}
	\en
with a density $n(\psi)$ and temperature $T(\psi)$ that are constant on each flux surface $\psi$.\footnote{In the limit of zero gyroradius, it is possible to express the Maxwellian in this way as a function of motion invariants. However, any such function will in general deviate from the Maxwellian to first order in gyroradius since particle orbits do not exactly follow field lines. A function depending only on constants of the motion cannot, in most cases, be exactly Maxwellian, and it is this discrepancy that drives neoclassical transport. For the present discussion, it is unimportant.} We shall assume that the distribution function evolves on a time scale longer than the bounce time for trapped particles, so that $\mu$ and $J$ remain constant for each such particle and the distribution function is independent of the position $l$ along $\bf B$. Our main task is to find the ground state as described by some distribution function $f_0(\psi, \alpha, \mu, J)$ of the trapped particles. As shown in I, ground states can be written as functions of energy and the conserved quantities alone, i.e., it must be possible to write $f_0$ as
	$$ f_0(\psi, \alpha, \mu, J) = F[\epsilon(\psi, \alpha, \mu, J), \mu, J] $$
for some function $F$. 

Before proceeding to find this function, it is useful ask the quesiton whether a Maxwellian flux function (\ref{Maxwellian flux function}) can represent a ground state. This can evidently only be the case if
	$$ \left( \frac{\p f}{\p \alpha} \right)_{\epsilon, \mu, J} = 0, $$
but for a Maxellian the left-hand side is equal to 
	$$ \left( \frac{\p}{\p \alpha} f_M (\epsilon,\psi)
	\right)_{\epsilon, \mu, J}
	= \frac{\p f_M}{\p \psi}  \left( \frac{\p \psi}{\p \alpha} \right)_{\epsilon, \mu, J}
	= - \frac{\p f_M}{\p \psi} 
	\left( \frac{\p J}{\p \alpha} \right)_{\epsilon, \mu, \psi} \bigg\slash
	\left( \frac{\p J}{\p \psi} \right)_{\epsilon, \mu, \alpha}. $$ 
We thus conclude that Eq.~(\ref{Maxwellian flux function}) can only represent a ground state if either the density and temperature are constant, so that $\p f_M/\p \psi = 0$, or the magnetic field is omnigenous, $\p J/\p \alpha = 0$. By definition, in an omnigenous field, $J$ can be written as a function of energy, magnetic moment and the flux-surface label $\psi$, without any dependence on $\alpha$, i.e., we can write $J=J(\epsilon,\mu,\psi)$. Since trapped-particle orbits precess on surfaces of constant $J$, omnigenous fields have the property that collisionless particle trajectories are radially well confined, making the neoclassical transport small [\cite{Hall-1975,Cary-1997,Helander-2014-a}]. It is interesting to note that there is thus a connection between low neoclassical transport -- an important goal of stellarator optimisation -- and small available energy, which could be beneficial for turbulence reduction. 

Of course, omnigeneity is merely a necessary, but not sufficient, condition for a ground state of the form (\ref{Maxwellian flux function}). The further condition that $\p F / \p \epsilon \le 0$ was shown in I to be equivalent to 
	\bn \frac{\omega_\ast^T}{\omega_\alpha} \le 1, 
	\label{ground-state condition}
	\en
where we have written
	$$ \omega_\ast = \frac{T}{q} \frac{d \ln n}{d \psi}, $$
  \bn \omega_\ast^T = \omega_\ast \left[1 + \eta \left( \frac{\epsilon}{T} - \frac{3}{2} \right) \right], 
	\label{omega-ast}
	\en
and $\eta = (d \ln T / d \psi) / (d \ln n / d \psi)$ as in I but now denote the charge by $q$. We have also introduced the bounce-averaged precession frequency for trapped particles,
	\bn \omega_\alpha = \frac{1}{\tau_b} \int_{l_1}^{l_2} ({\bf v}_d \cdot \nabla \alpha) \frac{dl}{v_\|} 
	= \frac{1}{q} \left( \frac{\p \epsilon}{\p \psi} \right)_{\mu, J, \alpha}
	\label{precession freq}
	\en
where ${\bf v}_d$ denotes the drift velocity and 
	$$ \tau_b = \int_{l_1}^{l_2} \frac{dl}{|v_\| |} = \left( \frac{\p J}{\p \epsilon} \right)_{\psi, \alpha, \mu} $$
the bounce time between two consecutive bounce points, which have been denoted by $l_{1,2}$. The precession frequency can also be written as
	$$ \omega_\alpha = -\frac{1}{q} \left( \frac{\p J}{\p \psi} \right)_{\epsilon, \mu, \alpha}
	\bigg\slash \left( \frac{\p J}{\p \epsilon} \right)_{\mu, \psi, \alpha}, $$
and is thus seen to be independent of $\alpha$ in an omnigenous field, where $\p J / \p \alpha = 0$. 

\subsection{Equations for the ground state}

We now turn to the calculation of the avaiable energy when the distribution function (\ref{Maxwellian flux function}) is not a ground state, i.e., when the condition (\ref{ground-state condition}) is not satisfied. For simplicity, we only consider omnigenous magnetic fields, such as that of a tokamak or well-optimised stellarator. 

When certain quantities $\bf y$ such as $\mu$ and $J$ are conserved, it is useful to employ these as (some of the) phase-space coordinates. We thus write the latter as ${\bf x} = ({\bf y}, {\bf z})$ and recall from I that the minimum-energy state is given by a simple generalisation of (\ref{Ground-state eq}),
	\bn H\left[ F(w,{\bf y}), {\bf y} \right] = \Omega(w, {\bf y}). 
	\label{ground state with mu}
	\en
where 
	$$ H(\phi, {\bf y}) = \int \Theta[f({\bf y}, {\bf z}) - \phi] \sqrt{g} \; d{\bf z}, $$
	$$ \Omega(w,{\bf y}) = 
	\int \Theta\left[w - \epsilon({\bf y}, {\bf z}) \right] \sqrt{g} \; d{\bf z}. $$

In the present case, we have two adiabatic invariants for the electrons, ${\bf y} = (\mu, J)$, and the phase-space volume element is (see I)
	\bn d\Gamma = d{\bf r} d{\bf v} = \frac{4 \pi dl }{m^2 |v_\| | \tau_b} d\mu dJ d\psi d\alpha. 
	\label{volume element}
	\en
We thus have the relations
	$$ H(\phi,\mu, J) = \frac{4 \pi}{m^2} \oint  d\alpha \int \Theta\left(f - \phi \right) d\psi, $$
	$$ \Omega(w,\mu,J) = \frac{4 \pi}{m^2} \oint  d\alpha
	\int \Theta\left[w - \epsilon(\psi,\alpha,\mu,J) \right] d\psi, $$
which together with Eq.~(\ref{ground state with mu}) determine the ground state. Here and in the following, integrals over $\alpha$ are to be taken over all field lines on the flux surface, $0 \le \alpha \le 2 \pi$, and we have used the fact that $f$ does not vary much along the field line, i.e., $f$ is nearly independent of $l$ at fixed $(\psi, \alpha, \mu, J)$. 

Rather than solving Eq.~(\ref{ground state with mu}) directly, we consider its derivative [\cite{Dodin-2005}],
	\bn \left( \frac{\p F}{\p w} \right)_{\mu,J} 
	= \frac{\Omega_1(w, \mu, J)}{H_1\left[F(w,\mu, J), \mu, J \right]}, 
	\label{Dodin-Fisch}
	\en
where the subscript $1$ denotes a derivative with respect to the first argument of the function in question. Thus
	$$ \Omega_1(w, \mu, J) = 
	\oint d\alpha  \int \delta \left[ w - \epsilon(\psi, \alpha, \mu, J) \right] \frac{4 \pi}{m^2} d\psi 
	= \frac{4 \pi}{m^2} \oint \frac{d\alpha}{|q \omega_\alpha |}, $$
where we have used Eq.~(\ref{precession freq}) and 
	$$ \int \delta \left[ w - \epsilon(\psi, \alpha, \mu, J) \right] d\psi = \frac{1}{| \p \epsilon / \p \psi |}. $$
For the denominator of (\ref{Dodin-Fisch}) we need
		$$ H_1\left(\phi, \mu, J \right) = 
		- \oint d\alpha  \int \delta \left[ f_M(\psi, \alpha, \mu, J) - \phi \right] \frac{4 \pi}{m^2} d\psi 
	= - \frac{4 \pi}{m^2} \oint
	\left| \left( \frac{\p f_M}{\p \psi} \right)_{\alpha, \mu, J} \right|^{-1} d\alpha , $$
which is equal to 
	$$ H_1\left(\phi, \mu, J \right) = \oint
	\frac{4 \pi T d\alpha}{m^2 \left| q \left(\omega_\ast^T - \omega_\alpha \right) \right| f_M }, $$
where we have used
		$$ \left( \frac{\p f_M}{\p \psi} \right)_{\alpha, \mu, J} 
		= \frac{q f_{M0}}{T} \left(\omega_\ast^T - \omega_\alpha \right), $$ 
Our equation (\ref{Dodin-Fisch}) for the ground state thus becomes
	$$ \left( \frac{\p F}{\p w} \right)_{\mu,J} 
	= - \frac{f_M}{T} \left| \frac{\omega_\ast^T}{\omega_\alpha} - 1 \right|. $$
	
\subsection{Available energy}

We can solve this equation in the neighbourhood of a flux surface, $\psi_0 - \Delta \psi < \psi < \psi_0 + \Delta \psi$, by expanding in $\psi - \psi_0$,
	$$ \epsilon(\mu,J,\psi,\alpha) = \epsilon_0(\mu, J) + \frac{\p \epsilon}{\p \psi} (\psi - \psi_0) + \cdots, $$
	$$ f_M(\mu,J,\psi,\alpha) = f_{M0} + \frac{\p f_M}{\p \psi} (\psi - \psi_0) +  \cdots, $$
	$$ F(\mu,J,\psi,\alpha) = F_0 + \frac{\p F}{\p \psi} (\psi - \psi_0) +  \cdots, $$
where all derivatives and all quantities with a subscript $0$ are to be evaluated at $\psi = \psi_0$, and we recall Eq.~(\ref{precession freq}) and 
	$$ \frac{\p F}{\p \psi} = -\frac{q f_{M0}}{T} \left| \omega_\ast^T-  \omega_\alpha \right|. $$ 
Because the volume element in phase space is given by Eq.~(\ref{volume element}) and the total number of particles having any given values of $\mu$ and $J$ must be the same in the initial state and the ground state, we have
	$$ \int_{\psi_0 - \Delta \psi}^{\psi_0 + \Delta \psi} (f_M - F) d\psi
	= \int_{\psi_0 - \Delta \psi}^{\psi_0 + \Delta \psi} \left[f_{M0} - F_0
	+ \frac{1}{2} \left( \frac{\p^2 f_M}{\p \psi^2} - \frac{\p^2 F}{\p \psi^2} \right) (\psi - \psi_0)^2 \right] d\psi
	= 0. $$
The available energy can therefore to leading order be calculated from
	$$ A = \int \epsilon (f_M - F) d\Gamma $$
as follows: 
	$$ A \simeq \int \left[ \epsilon_0 + q \omega_\alpha (\psi - \psi_0) \right] 
	\left[ f_{M0} - F_0 + \frac{q f_{M0}}{T} \left(\omega_\ast^T - \omega_\alpha 
	- \omega_\alpha \left| \frac{\omega_\ast^T}{\omega_\alpha} - 1 \right| \right)  (\psi - \psi_0) \right. $$
	$$ \left. + \frac{1}{2} \left( \frac{\p^2 f_M}{\p \psi^2} - \frac{\p^2 F}{\p \psi^2} \right) (\psi - \psi_0)^2 \right]d\Gamma $$
	$$ = \frac{q^2}{T} \int f_{M0} \omega_\alpha^2 \left(\frac{\omega_\ast^T}{\omega_\alpha} - 1 
	- \left| \frac{\omega_\ast^T}{\omega_\alpha} - 1 \right| \right) (\psi - \psi_0)^2 d\Gamma, $$
and the result becomes
	\bn A = \frac{16\pi q^2 (\Delta \psi)^3}{3 m^2 T} \int f_{M0} \omega_\alpha^2 
	r \left(  \frac{\omega_\ast^T}{\omega_\alpha} - 1 \right) d\mu dJ d\alpha, 
	\label{A with constant mu and J}
	\en
where $ r(x) = x \Theta(x) $. It can also be written as
	\bn A = \frac{4 q^2 (\Delta \psi)^3}{3T} U'(\psi) \int_{\rm trapped} f_{M0} \omega_\alpha^2 
	r \left(  \frac{\omega_\ast^T}{\omega_\alpha} - 1 \right) d{\bf v}, 
	\label{omnigenous A}
	\en
where $U(\psi)$ denotes the volume enclosed by the flux surface $\psi$. 

This expression for the maximum energy that can be extracted from electrons by fluctuations satisfying the ordering (\ref{ordering}) is one of the principal results of this paper. In order to explore how it scales with plasma parameters and magnetic-field geometry, we note that
	$$ \omega_\alpha \sim \frac{T}{m\Omega a R}, $$
	$$ \omega_\ast^T \sim \frac{T}{e a L_\perp B}, $$
where $a$ denotes the minor radius of the flux surface in question, $R$ the radius of curvature of the magnetic field lines, and $L_\perp$ the length scale of the pressure profile, which is assumed to be shorter than $R$. The volume enclosed by the flux surface $\psi$ is 
	$$ U(\psi) \sim \frac{R \psi}{B}, $$
and that of the region between $\psi - \Delta \psi$ and $\psi + \Delta \psi$
	$$ V \sim a R \Delta r, $$
where $\Delta \psi \sim aB \Delta r$. Within factors of order unity, the available energy (\ref{omnigenous A}) is thus given by
	$$ \frac{A}{V} \sim f_t nT \; \frac{(\Delta r)^2}{R L_\perp}, $$
where $f_t$ denotes the fraction of electrons that are trapped in regions of average unfavourable curvature. 

The result (\ref{omnigenous A}) for the available energy of electrons conserving $\mu$ and $J$ is always smaller than the corresponding energy without these constraints. The latter was calculated in I and was found to be equal to 
	$$ \frac{A}{V} = nT \lang \frac{1}{2} \left( \frac{\delta n}{n} \right)^2 
	+ \frac{3}{4} \left( \frac{\delta T}{T} \right)^2 \rang \sim nT \; \frac{(\Delta r)^2}{L_\perp^2}$$
for a Maxwellian plasma with small density and temperature variations, $\delta n \ll n$ and $\delta T \ll T$, respectively.

\section{Fixed density profile}

In many numerical simulations of plasma instabilities and turbulence, the electrons are assumed to be `adiabatic', meaning that the electron density fluctuations $\delta n_e$ are proportional to those of the electrostatic potential $\delta \Phi$ according to the linearised Boltzmann relation
	$$ \frac{\delta n_e}{n_e} = \frac{e \delta \Phi}{T_e}, $$
where $-e$ denotes the electron charge and $T_e$ the electron temperature. (In configurations with closed magnetic field lines or magnetic flux surfaces, an appropriate average of $\delta \Phi$ should be subtracted from the right-hand side.) Because the ${\bf E} \times {\bf B}$-drift is then out of phase with the density fluctuations, there is no electron particle flux across the magnetic field and the density profile must remain constant. 

In a plasma, as opposed to a numerical simulation, the assumption of `adiabatic' (or, better, Boltzmann-distributed) electrons is valid if the ordering (\ref{ordering}) holds and, additionally, no (or sufficiently few) electrons are trapped in magnetic wells. In this ordering, the transit frequency of the circulating electrons substantially exceeds the fluctuation frequency. These electrons thus travel many times around the torus on the time scale of one fluctuation, and therefore experience little transport. Mathematically, this conclusion follows, for instance, from a familiar argument for the confinement of circulating particles in a stellarator in the face of cross-field drifts [\cite{Helander-2014-a}]. The drift velocity, including the ${\bf E} \times {\bf B}$ drift, is
	$$ {\bf v}_{da} = \frac{v_\|}{B} \nabla \times \left( \frac{ v_\| \bf B}{\Omega_a} \right), $$
where the derivative is taken at constant $\mu_a$ and $H_a$. To first order in the smallness of the drift, the net displacement across the flux surface is thus
	$$ \Delta \psi = \int ({\bf v}_{da} \cdot \nabla \psi) \frac{dl}{v_\|} = 
	\int \nabla \cdot \left( \frac{v_\| {\bf B} \times \nabla \psi}{\Omega_a} \right) \frac{dl}{B}, $$
where the integral is taken along the field line. If the latter wraps many times around the torus, the integral becomes proportional to the flux-surface average of $\nabla \cdot [(v_\| / \Omega_a) {\bf B} \times \nabla \psi ]$, which vanishes since the vector within the divergence is perpendicular to $\nabla \psi$. An exception occurs close to surfaces on which the rotational transform (winding number of the magnetic field lines) is a rational number. In such regions, it takes many turns around the torus before a field line has come close to most points on the magnetic surface, and the electrons therefore do not respond ``adiabatically'' despite the ordering (\ref{ordering}) [\cite{Hallatschek-2005,Dominski-2015}].

\subsection{Ground state}

Suppose we ask for the minimum energy state of the ions in a plasma with fixed density profile. It will in general not be given by Eq.~(\ref{ground state}), because the latter does not account for the constraint that the density profile cannot change. Equation (\ref{ground state}) predicts a state of constant density throughout the plasma, which is not attainable unless the initial state had constant density. Instead, the ground state must be determined by minimising the energy subject to both the constraint (\ref{H}) and the requirement that the density profile is given,
	$$ \int f_0({\bf r}, {\bf v}) \; d{\bf v} = n({\bf r}) = \rm fixed. 
	$$
The latter can be accounted for by an additional set of Lagrange multipliers $\kappa({\bf r})$, so that instead of Eq.~(\ref{W}) we seek to minimise
	$$ U[f_0,\kappa,\lambda] = W[f_0,\lambda] + \int \kappa({\bf r}) d{\bf r}
	\left[ \int f_0({\bf r}, {\bf v}) \; d{\bf v} - n({\bf r}) \right]. $$
The variation with respect to $\kappa({\bf r})$ now ensures that the density profile is equal to $n({\bf r})$, and the variation with respect to $f_0$ now gives
	$$ \delta U = \int \delta f_0 \left[ \epsilon + \lambda(f_0) + \kappa({\bf r}) \right] \sqrt{g} \; d{\bf x} 
	= 0, $$
so that $\lambda(f_0) = - \epsilon - \kappa({\bf r})$. The ground-state distribution function must therefore be a function of the form
	\bn f_0({\bf r}, {\bf v}) = F[\epsilon({\bf r}, {\bf v}) + \kappa({\bf r})],
	\label{density-contrained ground state}
	\en
rather than (\ref{ground state}), and the density contraint becomes
	\bn \int F[\epsilon({\bf r}, {\bf v}) + \kappa({\bf r})] d{\bf v} = n({\bf r})
	\label{density}
	\en
	
If we take $\epsilon = mv^2/2$, we can conclude a particular example of a ground state is an isothermal Maxwellian,
		$$ f = n({\bf r}) \left( \frac{m}{2 \pi T} \right)^{3/2} e^{- \frac{m v^2}{2 T}}, $$
where $\nabla T = 0$. It follows that such a distribution function is linearly and nonlinearly stable to fluctuations that do not disturb the density profile. This conclusion is not surprising but has perhaps not been proved before. Moreover, it follows that any Maxwellian that is {\em not} isothermal is not in a ground state, even if it happens to be linearly stable. For instance, a plasma with $0 < \eta = |\nabla \ln T| / | \nabla \ln n | < 2/3$ is linearly stable to ion-temperature-gradient modes with adiabatic electrons [\cite{Hazeltine-1998}] but has non-zero available energy and could thus potentially support subcritical turbulence [\cite{Newton-2010,Plunk-2013,van-Wyk-2016,McMillan-2018}].

\subsection{Available energy}

In order to calculate the available energy of such a plasma, we write the initial distribution function as
	$$ f({\bf x},0) = M({\bf r}) e^{- \frac{m v^2}{2 T({\bf r})}} $$
and explore the consequences of Liouville's theorem (\ref{H}). Since $f({\bf x},0) > \phi$ if and only if
	$$ v^2 < \frac{2T}{m} \ln \frac{M}{\phi}, $$
which defines a sphere in velocity space of volume
	$$ \frac{4 \pi}{3} \left( \frac{2 T}{m} \ln \frac{M}{\phi} \right)^{3/2}, $$
we have
	$$ H[f({\bf x},0),\phi] = 
	\frac{4 \pi V}{3} \lang \left( \frac{2 T}{m} \ln \frac{M}{\phi} \right)^{3/2} 
	\; \Theta(M-\phi) \rang $$
and
	$$ H[f_0({\bf x}),\phi] = \int \Theta \left[ F \left(\epsilon({\bf v}) 
	+ \kappa({\bf r}) \right) - \phi \right]
	d{\bf x}.
	$$
In these equations, ${\bf x} = ({\bf r}, {\bf v})$, $V$ denotes the plasma volume and angular brackets a volume average. If we denote the inverse of $F$ by $w$, so that $F[w(\phi)] = \phi$, then $F \left(\epsilon({\bf v}) 
	+ \kappa({\bf r}) \right) = \phi$ when 
		$$ v = \sqrt{ \frac{2}{m} \left[ w(\phi) - \kappa({\bf r}) \right]}, $$
and we conclude that the phase-space volume in which $f_0({\bf x})$ exceeds $\phi$ is equal to
	$$ H[f_0({\bf x}),\phi] = \frac{4 \pi V}{3} 
	\lang \left[ \frac{2}{m} ( w(\phi) - \kappa({\bf r}) ) \right]^{3/2}_\Theta \rang, $$
where we have introduced the notation
	$$ \left[ y \right]^{3/2}_\Theta = y^{3/2} \Theta(y). $$
Equation (\ref{H}) thus becomes
	\bn 	\lang \left[ w(\phi) - \kappa({\bf r}) \right]^{3/2}_\Theta \rang = 
	\lang \left[ T({\bf r}) \ln \frac{M({\bf r})}{\phi} \right]^{3/2}_\Theta \rang,
	\label{nonlinear ground state}
	\en
and constitutes an integral equation for the function $w(\phi)$ -- or, alternatively, for its inverse $F(w)$. [The corresponding equation (3.14) in I is obtained by setting $\kappa({\bf r}) = 0$.] We thus have two nonlinear integral equations, (\ref{density}) and (\ref{nonlinear ground state}), for the two unknown functions $\kappa$ and $F$. 

We cannot construct the general solution explicitly, but if the density and temperature of the initial state only vary slightly, it is possible to make progress. We then write 
	$$ n({\bf r}) = \lang n \rang [1 + \nu({\bf r}) ], $$
	$$ T({\bf r}) = \lang T \rang [1 + \tau({\bf r}) ], $$
with $\nu \sim \tau \ll 1$, so that the initial energy density is
	$$ E_M = \frac{3}{2} \lang nT \rang = \frac{3}{2} \lang n \rang \lang T \rang \lang 1 + \nu \tau \rang, $$
and seek an approximate solution of the form
	\bn F(w) = \bar M e^{-w / \lang T \rang + h(w)}, 
	\label{Ansatz}
	\en
where $\bar M = \lang n \rang (m / 2 \pi \lang T \rang)^{3/2}$ and $h \ll 1$. Before trying to solve Eq.~(\ref{nonlinear ground state}), it is helpful to consider the density constraint (\ref{density}), which becomes
	$$ \lang n \rang [1 + \nu({\bf r}) ] = \bar M e^{-\kappa({\bf r}) / \lang T \rang}
	\int \exp \left[-\frac{\epsilon({\bf v})}{\lang T \rang}  + h(\epsilon({\bf v}) + \kappa({\bf r}))\right]
	d{\bf v}. $$
Anticipating that $h$ will turn out to be of order $O(\nu^2)$ and $\kappa$ of order $O(\nu)$, we can approximate the right-hand side by
	$$ \bar M e^{-\kappa({\bf r}) / \lang T \rang}
	\int \exp \left[-\frac{\epsilon({\bf v})}{\lang T \rang}  + h(\epsilon({\bf v}))\right]
	d{\bf v}, $$
where only $\kappa({\bf r})$ depends on ${\bf r}$. It follows that
	$$ \frac{\kappa}{\lang T \rang} = - \nu + \frac{\nu^2}{2} + O\left(\nu^3 \right). $$
The ground-state equation (\ref{nonlinear ground state}) now becomes, to the requisite accuracy, 
	$$ \lang \left(x + \nu - \frac{\nu^2}{2} \right)^{3/2} \rang
	= \lang (1 + \tau )^{3/2} \left( x + \ln \frac{1 + \nu}{(1 + \tau)^{3/2}} - h \right)^{3/2} \rang, $$
where we have written $x = w/\lang T \rang$. Expanding both sides to second order in $\nu\sim \tau \ll 1$ and solving for $h$ gives after some algebra
	$$ h(w) = \frac{\lang \tau^2 \rang}{4}x  + \frac{3}{2} \lang \nu\tau - \tau^2 \rang
	+ \frac{3}{4x} \lang \frac{3 \tau^2}{4} - \nu \tau \rang, $$
confirming our earlier assumption that $h = O(\nu^2)$. The ground-state distribution function is thus given by
	$$ F(\epsilon + \kappa) = \bar M e^{- \epsilon / \lang T \rang} \left[ 1 + \nu + h(\epsilon) \right], $$
and the associated energy per unit volume
	$$ E_0 = \lang \int \epsilon F(\epsilon + \kappa) d{\bf v} \rang 
	= \frac{3}{2} \lang n \rang \lang T \rang \lang \nu \tau - \frac{\tau^2}{2} \rang. $$
The available energy is thus
	\bn \frac{A}{V} = E_M - E_0 = E_0 \lang \frac{\tau^2}{2} \rang. 
	\label{new Vlasov A}
	\en

In contrast, if the density profile is free to evolve, the available energy becomes
	\bn \frac{A}{V}  	
	= E_0 \lang \frac{\nu^2}{3} + \frac{\tau^2}{2} \rang, 
	\label{Vlasov A}
	\en
as originally found in I without explicitly calculating the distribution function. Thus, as one might naively expect, in a plasma with fixed density profile only the temperature variations contribute to the free energy and their contribution is the same as if the density profile were free to evolve. Lest this conclusion be thought to be self-evident, it should be noted that it does not actually hold unless the density and temperature variations are small. This fact will be illustrated by a simple example in the next section.

\section{Mixing of two different plasmas}

We now demonstrate that the amount of available energy from temperature variations within the plasma in general depends on the density profile even if the latter is held fixed. To this end, consider a Maxwellian plasma occupying a volume consisting of two subvolumes, $V = V_1 + V_2$, where the density and temperature are constant in each subvolume, so that $n({\bf r}) = n_j$ and $T({\bf r}) = T_j$ for ${\bf r} \in V_j$, $j = 1,2$. Moreover, we assume that $T_2 / T_1 \ll 1$ but $n_1 V_1 \sim n_2 V_2$ and anticipate that the ground state is one where the hot plasma in $V_1$ and the cold plasma in $V_2$ have mixed in such a way that the total energy is reduced. The initial plasma energy is
	$$ E = \frac{3}{2} (n_1 T_1 V_1 + n_2 T_2 V_2) \simeq  \frac{3 n_1 T_1 V_1}{2}, $$
and will be compared with the minimum attainable energy assuming either that the density is free to evolve or that it is fixed in each of the two subvolumes. 

\subsection{Unconstrained density profile}

If the density profile is free to evolve, the equation (\ref{Ground-state eq}) for the ground state becomes
	$$ \epsilon^{3/2} = 
	 \frac{V_1}{V} \left[ T_1 \ln \frac{M_1}{F(\epsilon)} \right]^{3/2}_\Theta
	+ \frac{V_2}{V} \left[ T_2 \ln \frac{M_2}{F(\epsilon)} \right]^{3/2}_\Theta, $$
where $M_j = n_j (m/2\pi T_j)^{3/2}$. This equation follows immediately from Eq.~(3.14) in I and is similar to Eq.~(\ref{nonlinear ground state}) with $\kappa = 0$ in the present paper. In the limit $T_2 / T_1 \rightarrow 0$, the first term on the right dominates in most of velocity space, where $\epsilon \sim T_1 \gg T_2$, whereas the second term dominates for energies comparable to $T_2$. The solution is thus
	$$ F(\epsilon) = \left\{ \begin{array}{ll}  M_1 e^{-\epsilon / \tilde T_1}, & \epsilon \sim T_1 \\                  M_2 e^{-\epsilon / \tilde T_2},  & \epsilon \sim T_2     
						\end{array}       \right.  $$
where $\tilde T_j = T_j (V_j/V)^{2/3}$.\footnote{These asymptotic relations determine the solution in most of velocity space and do so sufficiently accurately for our purposes but fail in the energy range $T_2 \ll \epsilon \ll T_1$, where a smooth transition occurs.} This solution describes a bi-Maxwellian distribution such that the density of each component is constant throughout the entire plasma volume $V$. The energy associated with this distribution function is
	$$ E_0 = V \int \epsilon F d{\bf v} \simeq \frac{3 n_1 T_1 V}{2} \left( \frac{V_1}{V} \right)^{5/3}, $$
and the available energy thus becomes
	\bn A = E-E_0 = E \left[1 - \left( \frac{V_1}{V} \right)^{2/3} \right]. 
	\label{two-volume A}
	\en
This is also the energy that is released if the hot plasma expands adiabatically (with pressure $p \propto V^{-5/3}$) from $V_1$ into the greater volume $V$. 

Note that the result (\ref{two-volume A}) does not lead to any definite prediction of what will actually happen to the plasma, only that it is possible that thermal energy can be converted into kinetic energy. It should come as no surprise that, even in the collisionless case under consideration, the amount of convertible thermal energy is bounded by classical thermodynamics.

\subsection{Fixed density profile}

If the density is constrained to remain equal to its initial value in each of the two subvolumes, the distribution function of the ground state no longer depends on energy alone but is a function of the form (\ref{density-contrained ground state}), where the function $\kappa({\bf r})$ enforcing the density constraint can be anticipated to be constant in each of the two subvolumes, $\kappa({\bf r}) = \kappa_j$ for ${\bf r} \in V_j$. The ground-state equation (\ref{nonlinear ground state}) then becomes
	$$ 	V_1 \left[ w - \kappa_1 \right]_\Theta^{3/2}
	+ V_2 \left[ w - \kappa_2 \right]_\Theta^{3/2} = 
	V_1 \left[ T_1 \ln \frac{M_1}{\phi} \right]_\Theta^{3/2}
	+ V_2 \left[ T_2 \ln \frac{M_2}{\phi} \right]_\Theta^{3/2}.
	$$
It does not seem possible to write down an explicit solution to this algebraic equation for $w(\phi)$ except in the special case that the density vanishes in one of the two regions, $M_2 \rightarrow 0$, say. The second term on the right then vanishes and $\kappa_2 \rightarrow \infty$, so that the solution becomes
	$$ F[\epsilon + \kappa({\bf r})] = \left\{ \begin{array}{ll}  M_1 e^{-\epsilon / T_1}, & {\bf r} \in V_1 \\  
			0, & {\bf r} \in V_2     
						\end{array}       \right.  $$
As one would expect from the fact that the density cannot change, the particles cannot move from one subvolume to the other, the distribution function remains equal to the initial condition, and the available energy thus vanishes, $A=0$. We conclude that the contribution to the available energy from temperature variations within the plasma in general depends on the density profile and on whether this profile is free to evolve or is constrained to remain constant. 

\section{Two species with different constraints}

We finally turn to the case discussed in the Introduction of a plasma where the magnetic moment is conserved for both the ions and the electrons, but the second adiabatic invariant only for the latter. The central question is whether it is possible to confine a plasma in a Maxwellian ground state for both species with respect to perturbations that conserve $\mu_i$, $\mu_e$ and $J_e$, and, additionally, satsify the requirement of quasineutrality. Following the method developed in Sections 2-4, possible ground states can be found be minimising the energy functional augmented by terms with Lagrange multipliers ensuring that the appropriate constraints are satisfied. 

Some care is required because different constraints pertain to the trapped and passing electrons. As discussed at the beginning of Section 4, the latter experience little transport when the ordering (\ref{ordering}) is adopted. Since $\psi$ is is thus conserved for each passing electron, it is advantageous to use the phase-space coordinates $(\mu,\epsilon, \psi, \alpha, l)$ for these particles. Because of rapid streaming along the field lines, the distribution function $f_{ec}$ of circulating electrons is approximately independent of $\alpha$ and $l$, and the local density of these particles becomes
	$$ n_{ec} = \frac{4 \pi B}{m_e^2} \int_{\mu B < \epsilon} f_{ec} (\mu, \epsilon, \psi) \frac{d\mu d\epsilon}{|v_\||}, $$
if, for simplicity, the numbers of co- and counter-moving particles are taken to be equal. A similar expression holds for the ion density if we use analogous phase-space coordinates for the ions. For trapped electrons, however, it is more appropriate to use the coordinates $(\mu, J, \psi, \alpha, l)$, so that the volume element in phase space is given by Eq.~(\ref{volume element}) and the local density becomes
$$ n_{et}({\bf r}_0) = \int f_{et} \delta({\bf r} - {\bf r}_0) d\Gamma $$
	$$ = \frac{4 \pi}{m_e^2} \int f_{et}(\psi, \alpha, \mu, J) \left| \frac{\p(\psi, \alpha, l)}{\p {\bf r}} \right|
	\delta(\psi - \psi_0) \delta(\alpha - \alpha_0) \delta(l -l_0) \frac{dl}{\tau_b|v_\| |} d\mu dJ d\psi d\alpha. $$
where subscripts $0$ refer to the point ${\bf r}_0$ and $f_{et}$ denotes the distribution function of trapped electrons. Since the Jacobian is                                                       
		$$ \left| \frac{\p(\psi, \alpha, l)}{\p {\bf r}} \right| = B, $$
the density of trapped electrons becomes
	$$ n_{et}({\bf r}_0) = \frac{4 \pi B_0}{m_e^2} 
	\int f_{et}(\psi_0, \alpha_0, \mu, J) \left| v_\| \tau_b \right|_0^{-1} d\mu dJ. $$
The ambipolarity condition is thus $Q[f_i, f_{ec}, f_{et}, \psi, \alpha, l] = 0$, where
		$$ Q[f_i, f_{ec}, f_{et}, \psi, \alpha, l] = 
		\frac{4 \pi B}{m_i^2} \int_{T \cup C} f_i \frac{d \mu d\epsilon}{|v_\||} -
		\frac{4 \pi B}{m_e^2} \left( \int_{C} f_{ec} \frac{d \mu d\epsilon}{|v_\||} + 
	\int_{T} f_{et}  \frac{d\mu dJ}{ \left| v_\| \tau_b \right|} \right). $$
Here $C$ refers to the circulating region of velocity space,
	$$ 0 \le \mu < \frac{\epsilon}{B_{\rm max}}, $$
$T$ to the trapped region,
	$$ \frac{\epsilon}{B_{\rm max}} \le \mu \le \frac{\epsilon}{B}, $$
and $B_{\rm max}(\psi)$ denotes the maximum magnetic field strength on the surface labelled by $\psi$. Similarly, the energy functional is
	$$ E[f_i, f_{ec}, f_{et}] = 4\pi \int d\psi d \alpha dl \left( 
	\int_{T \cup C} \epsilon f_i \frac{d \mu d\epsilon}{m_i^2 |v_\||} \right. $$
	$$ \left.
	+ \int_C \epsilon f_{ec} \frac{d \mu d\epsilon}{m_e^2 |v_\||} + 
	\int_T \epsilon f_{et}  \frac{d\mu dJ}{ m_e^2 \left| v_\| \tau_b \right|} \right), $$
and possible ground states can be found be seeking the minimum of the functional 
	$$ W_1[f_e, f_i, \lambda_e, \lambda_i, \nu] = E[f_i, f_{ec}, f_{et}] 
	+ \int \nu(\psi,\alpha,l) Q[f_i, f_{ec}, f_{et}, \psi, \alpha, l] d\psi d\alpha dl $$
	$$ + \frac{4\pi}{m_i^2} \int_{T \cup C} \lambda_i(\phi,\mu) d\phi d\mu
	\int \left[ \Theta(f_i - \phi) - H_i(\phi,\mu) \right] \frac{d\epsilon}{| v_\| |} d \psi d\alpha dl$$
		$$ + \frac{4\pi}{m_e^2} \int_C  \lambda_{ec}(\phi,\mu,\psi) d\phi d\mu d\psi
	\int\left[ \Theta(f_{ec} - \phi) - H_{ec}(\phi,\mu, \psi) \right] \frac{d\epsilon}{|v_\| |} d\alpha dl. $$
	$$ + \frac{4\pi}{m_e^2} \int_T \lambda_{et}(\phi,\mu,J) d\phi d\mu dJ
	\int \left[ \Theta(f_{et} - \phi) - H_{et}(\phi,\mu, J) \right] d \psi d\alpha. $$
In this expression, the first term on the right represents the total energy of the ions and electrons, and the next term, which contains the Lagrange multiplier $\nu(\psi,\alpha,l)$, enforces quasineutrality on each field line $(\psi,\alpha)$. The term containing $\lambda_i(\phi)$ ensures the conservation of $\mu$ for all ions, that containing $\lambda_{ec}(\phi, \mu, \psi)$ makes sure that $\mu$ and $\psi$ are conserved for circulating electrons, and the term with $\lambda_{et}(\phi,\mu,J)$ enforces conservation of $\mu$ and $J$ for trapped electrons. 

The variation with respect to $f_i$ gives
	$$ \int \delta f_i \left[ \epsilon + \nu + 
	\int \lambda_i(\phi, \mu) \delta (f_i - \phi) d\phi\right] \frac{d\mu_i d\epsilon}{| v_\| |} d\psi d\alpha dl = 0, $$
which implies  
	$$ \epsilon + \nu + \lambda_i(f_i, \mu) = 0. $$
It must therefore be possible to write $f_i$ as a function of the form
	\bn f_i = F_i[\epsilon + \nu(\psi,\alpha), \mu], 
	\label{fi}
	\en
where we have noted that rapid streaming along $\bf B$ implies that the distribution function must be independent of $l$, and so must therefore the function $\nu$. Similarly, the variation of $W_1$ with respect to the electron distribution function shows that the latter can be written as
	\bn 
	f_{ec} = F_{ec}[\epsilon - \nu(\psi,\alpha), \mu, \psi],
	\label{fec}
	\en
	\bn 
		f_{et} = F_{et}[\epsilon - \nu(\psi,\alpha), \mu, J]. 
	\label{fet}
	\en

\subsection{Maxwellian ground states}
	
Any possible ground states in plasmas with conservation properties derived from Eq.~(\ref{ordering}) must be of the form indicated by Eqs.~(\ref{fi})-(\ref{fet}), but the most important ones are of course those where the distribution functions of both species are Maxwellian. Equation (\ref{fi}) implies that this is only possible if the ions are isothermal, so that 
	$$ f_i = n(\psi) \left( \frac{m_i}{2 \pi T_i} \right)^{3/2} e^{-\epsilon / T_i}, $$
where $T_i$ is constant and we have taken $\nu(\psi,\alpha) = - T_i \ln n(\psi)$. For the electrons, however, the temperature may vary across magnetic flux surfaces,
	$$ f_e = n (\psi) \left( \frac{m_e}{2 \pi T_e(\psi)} \right)^{3/2} e^{-\epsilon / T_e(\psi)}. $$
In order that $f_{e}$ be of the form (\ref{fec}) and (\ref{fet}), we write $\kappa = \epsilon - \nu(\psi)$ and define
	$$ F_{ec}(\kappa, \mu, \psi) = N(\psi) 
		\left( \frac{m_e}{2 \pi T_e(\psi)} \right)^{3/2} e^{-\kappa / T_e(\psi)}, $$
	$$ F_{ec}(\kappa,\mu, J) = N[\psi(\kappa,\mu, J)] 
	\left( \frac{m_e}{2 \pi T_e[\psi(\kappa,\mu, J)]} \right)^{3/2} e^{-\kappa / T_e[\psi(\kappa,\mu, J)]}, $$
where the flux function $N(\psi)$ is related to the density by
	$$ n(\psi) = N(\psi) e^{\nu(\psi) / T_e(\psi)}. $$
	
This prescription suggests that it could be possible to confine a plasma in a Maxwellian ground state for both species with respect to perturbations that satisfy the ordering (\ref{ordering}) if the ions are isothermal. However, we must remember that Eqs.~(\ref{fi}) - (\ref{fet}) only provide necessary, and not sufficient, conditions on any ground states. In the case of a single species without conserved quantities, the corresponding condition is given by Eq.~(\ref{ground state}), but a ground state must also satisfy $dF/d\epsilon \le 0$ everywhere. To find the corresponding condition for the present two-species case, let us again consider what happens to the energy of the system when the particles in two neighbouring and equally large volume elements in phase space are interchanged, but we now allow such an interchange to occur both in the population of ions and in that of trapped electrons. 

As in Section 2, the energy released is equal to
	$$ dE_i = - d \epsilon_i df_i d\Gamma_i, $$
	$$ dE_e = -d \epsilon_e df_e d\Gamma_e, $$
where $d\Gamma_a$ denotes the phase-space volume of the elements, $d\epsilon_a$ their difference in energy and $df_a$ the difference in the distribution functions. In order that the interchanges be ambipolar, we must require
	\bn d \psi_i df_i d\Gamma_i = d \psi_e df_e d\Gamma_e, 
	\label{ambipolarity}
	\en
and conclude that the total amount of energy released is
	$$ dE = dE_i + dE_e = - d \epsilon_e df_e d\Gamma_e \left(1 + \frac{d\epsilon_i d\psi_e}{d\epsilon_e d \psi_i} \right). $$
For the trapped electrons, $d\psi_e$ and $d\epsilon_e$ are related by Eq.~(\ref{precession freq}),
	$$ d\epsilon_e = - e \omega_{\alpha e} d\psi_e, $$
which has to do with the conservation of $J_e$: it is in general impossible to change the energy of a trapped electrons without also moving it radially. We also have
	$$ df_e = \left( \frac{\p f_{et}}{\p \epsilon} \right)_{\mu, J} d\epsilon_e = 
	\frac{f_{et}}{T_e} \left( \frac{\omega_{\ast e}^T}{\omega_{\alpha e}} - 1 \right) d\epsilon_e, $$ 
and thus 
	$$ dE = - \frac{f_{et}}{T_e} \left( \frac{\omega_{\ast e}^T}{\omega_{\alpha e}} - 1 \right) \left( d\epsilon_e \right)^2
	\left(1 - \frac{1}{e \omega_\alpha} \frac{d \epsilon_i}{d \psi_i} \right). $$
The first term in the last bracket describes the energy change $dE_e$ in the electron channel and second one that in the ions. As found in I, the former is always positive if the condition (\ref{ground-state condition}) is satisfied, 
	$$ \frac{\omega_{\ast e}^T}{\omega_{\alpha e}} - 1 < 0, $$
but the total energy may nevertheless decrease. If more energy is released by the ions than taken up the electrons, so that $dE < 0$, the system could not have been in a ground state before the interchange. This is evidently the case if $d\epsilon_i$ and $d\psi_i$ are chosen in such a way that the condition
	\bn \frac{1}{e \omega_{\alpha e}} \frac{d \epsilon_i}{d \psi_i} > 1 
	\label{A>0}
	\en
is satisfied. Whether this is possible could, in principle, depend on the density gradient and the magnetic geometry. Since
	$$ df_i = \left( \frac{\p f_i}{\p \epsilon} \right)_\psi d \epsilon_i 
	+ \left( \frac{\p f_i}{\p \psi} \right)_{\epsilon} d\psi_i 
	= \frac{f_i}{T_i} \left( e \omega_{\ast i} d\psi_i - d \epsilon_i \right), $$
the ambipolarity condition (\ref{ambipolarity}) can be written as
	$$ \frac{f_i}{T_i} \left( e \omega_{\ast i} d\psi_i - d \epsilon_i \right) d\psi_i d\Gamma_i
	= \frac{e f_e}{T_e} \left( \frac{\omega_{\ast e}^T}{\omega_{\alpha e}} - 1 \right) d\epsilon_e d \psi_e d \Gamma_e, $$
and implies
	\bn \frac{1}{e \omega_{\alpha e}} \frac{d \epsilon_i}{d \psi_i} = \frac{\omega_{\ast i}}{\omega_{\alpha e}}
	+ \left( \frac{\omega_{\ast e}^T}{\omega_{\alpha e}} - 1 \right) 
	\left( \frac{d \psi_e}{d\psi_i} \right)^2 \frac{T_i f_e d\Gamma_e}{T_e f_i d\Gamma_i}. 
	\label{final criterion}
	\en

The question is whether this expression can be made to exceed unity by any choice of the coordinates and volumes of the phase-space volume elements that are to be interchanged. If so, the plasma is not in a ground state according to Eq.~(\ref{A>0}). Since $\omega_{\alpha e}$ is proportional to energy, the two terms on the right that have this quantity in the denominator can be made arbitrarily large by choosing the energy to be small.\footnote{An exception occurs if the density gradient vanishes, so that $\omega_{\ast e} = \omega_{\ast i} = 0$.} And since $\omega_{\ast e}$ and $\omega_{\ast i}$ have opposite signs, it is clear that Eq.~(\ref{final criterion}) will always exceed unity for some choice of paramenters. We thus conclude that non-trivial ground states do not exist in the present situation. In other words, it does not seem possible to confine a two-component plasma in a Maxwellian minimum-energy state with respect to fluctuations satisfying Eq.~(\ref{ordering}), unless the density and temperature gradients vanish. If the criterion (\ref{ground-state condition}) is satisfied for the electrons, their energy cannot be lowered, but it is apparently always possible to redistribute ions in such a way that the total energy decreases and quasineutrality is maintained. 

Finally, let us put this conclusion in the context of microinstabilities in stellarators and dipoles. As noted in I, the criterion (\ref{ground-state condition}) for the ground state of a single species can be met in two different ways. It holds if either the average magnetic curvature is favourable for all trapped particles and the temperature gradient is moderate,
	$$ \omega_\ast \omega_\alpha < 0, 
	$$
	$$ 0 < \eta < \frac{2}{3}, 
	$$
or, alternatively, if the curvature is {\em unfavourable} and the temperature gradient large enough, i.e., 
	\bn \omega_\ast \omega_\alpha > 0, 
	\label{dipole stability}
	\en
	$$ \eta > \frac{2}{3}. $$

The first one of these cases holds in a maximum-$J$ stellarator and implies that ordinary density-gradient-driven trapped-electron modes are linearly stable. \cite{Proll-2012} and \cite{Helander-2013} concluded that any collisionless instability must then be draw energy from the ions rather than the electrons, and \cite{Plunk-2017-b} showed that there is indeed a remnant ion-driven trapped-electron mode if $\eta_i > 0$. From our considerations of the available energy, this is not surprising, since it is apparently impossible to achieve a non-trivial ground state for both species simulataneously in the ordering (\ref{ordering}). On the other hand, if the electrons are in a ground state, they act to stabilise any ion-driven instability and one would expect that the available energy is relatively small. Gyrokinetic simulations of turbulence in approximately maximum-$J$ stellarator configurations do indeed suggest that the transport under these conditions is up to an order of magnitude smaller than in a typical tokamak [\cite{Helander-2015}]. 

As discussed in I, the second case (\ref{dipole stability}) can apply in the field of a magnetic dipole, which is then nonlinearly stable with respect to instabilities that conserve $\mu$ and $J$ for all species. This is of particular importance for electron-positron plasmas, where identical constraints apply to both species [\cite{Helander-2014}]. In electron-ion dipole plasmas, however, no such statement can be made concerning instabilities that do not conserve $J_i$, but low-frequency modes can be remarkably stable [\cite{Hasegawa-1990,Kesner-2002,Simakov-2002}].

\section{Conclusions}

The available energy of a plasma -- the amount of thermal energy that can be converted into linear and nonlinear fluctuations -- depends on what constraints limit the possible forms of plasma motion. In the present paper, this concept has been explored in the context of magnetically confined plasmas in local thermodynamic equilibrium. Specifically, an explicit formula (\ref{omnigenous A}) has been derived for the energy available from a species for which the first and second adiabatic invariants, $\mu$ and $J$, are conserved. This formula shows that the available energy is equal to a weighted average of $ \omega_\alpha^2 (\omega^T_\ast / \omega_\alpha - 1)$ over the subset of phase space where this quantity is positive. 
	
Frequently, different constraints apply to the electrons and the ions, but their combined motion must be such that quasineutrality is preserved. This condition adds another constraint and thus lowers the available energy. It also changes the functional form of the distribution function for the ``ground state'' of lowest accessible energy. In a plasma with fixed density profile, this distribution function no longer depends on energy alone, but on energy plus a function of position ensuring ambipolarity, see Eq.~(\ref{density-contrained ground state}). Since the density profile cannot relax, there is no available energy associated with the density gradient, and in a Vlasov plasma with small fluctuations this energy is given by Eq.~(\ref{new Vlasov A}) rather than Eq.~(\ref{Vlasov A}), which applies in a plasma where the density is free to evolve. Thus, small density variations do not affect the available energy of a plasma with fixed density profile. However, if the density variations are {\em not} small, then they {\em can} affect the amount of energy that can be extracted from temperature variations in the plasma, as demonstrated in Section 5.

Finally, we have noted that for some of the most important forms of turbulence in magnetically confined fusion plasmas, the magnetic moment $\mu$ is conserved for the ions whereas both $\mu$ and the parallel invariant $J$ are conserved for trapped electrons, while $\mu$ and $\psi$ are conserved for circulating electrons. This is, for instance, usually the case for ion-temperature-gradient driven modes and trapped-electron modes (TEMs), which are believed to cause much of the turbulence and transport observed in fusion experiments. In the collisionless, electrostatic limit, \cite{Proll-2012} and \cite{Helander-2013} derived the criterion (\ref{ground-state condition}) for the stability of TEMs taking energy from the electrons in the plasma. This condition came out of lengthy gyrokinetic calculations but can be simply understood in terms of the available energy. When this criterion is satisfied, any instability satisfying the orderings must derive its energy from the ions in the plasma, and \cite{Plunk-2017-b} identified a particular such instability. The considerations of available energy given in Section 6 suggest that it is indeed always possible to lower the total energy of a Maxwellian plasma with non-zero density or temperature gradients. 

Before closing, we note that all our results are, strictly speaking, only valid in zero-$\beta$, collisionless plasmas. Collisions violate Liouville's theorem and can enable the plasma to access lower-energy states than otherwise possible. Similarly, allowing for magnetic fluctuations can fundamentally change the stability properties of a plasma. However, experience from gyrokinetic simulations suggests that a small amount of collisions or plasma pressure does not substantially affect the turbulence. Insofar that the properties of the latter are correlated with the availalable energy, one can thus hope that this remains true beyond the strict realm of applicability of the calculations presented here.

\section*{Appendix: Available potential energy of the atmosphere}

In this Appendix, we establish the relation between our formalism and the results of \cite{Lorenz-1955}. Consider an atmosphere consisting of an ideal gas satisfying the equations
	$$ \frac{\p \rho}{\p t} + \nabla \cdot ( \rho {\bf v} ) = 0, $$
	$$ \frac{\p s}{\p t} + {\bf v} \cdot \nabla s = 0, $$
where $\rho$ denotes the density, ${\bf v}$ the flow velocity and $s = p/\rho^\gamma$ the specific entropy. Here $p$ is the pressure and $\gamma$ the adiabatic index (ratio of specific heats). The first one of these equations expresses conservation of mass and the second one that of entropy. Any function $f(s)$ of $s$ alone then satisfies the conservation law
	$$ \frac{d}{dt} \int \rho f(s) dV = 0, $$
where the volume integral is taken over the entire atmosphere and the normal component of $\rho {\bf v}$ is assumed to vanish on the boundary. Specifically, we may choose $f$ to be a Heaviside function and conclude that the quantity
	$$ H(\sigma) = \int \rho \Theta(s - \sigma) dV $$
does not change with time for any value of $\sigma \in [0,\infty)$. 
	
To find the ground state, we follow the recipe from Section 2 and minimise the energy functional
		$$ E = \int \left( \frac{p}{\gamma -1 } + \rho \phi \right) dV, $$
where $\phi = gz$ denotes the gravitational potential, subject to the contraint that $H(\sigma)$ is fixed. We are thus led to consider the functional
	$$ W[\rho,s,\lambda] = \int \left( u + \rho \phi \right) dV
	+ \int_0^\infty \lambda(\sigma) d\sigma \left( \int \rho \Theta(s-\sigma) dV - H(\sigma) \right), $$
where 
	$$ u(\rho,s) = \frac{s \rho^\gamma}{\gamma -1 } $$
and the Lagrange multipliers $\lambda(\sigma)$ ensure that $H(\sigma)$ is indeed conserved for each $\sigma \ge 0$. The variation of $W$ with respect to $\rho$ and $s$ gives
	$$ \phi + \frac{\p u}{\p \rho} + \int_0^\infty \lambda(\sigma) \Theta(s-\sigma) d\sigma = 0, 
	$$
	$$ \frac{\p u}{\p s} + \lambda(s) \rho = 0,
	$$
respectively. The second of these equations implies 
	\bn \lambda(s) = - \frac{\rho^{\gamma - 1}}{\gamma - 1}, 
	\label{s variation}
	\en
which, when substituted into the gradient of the first equation,
	$$ \nabla \phi + \frac{\gamma}{\gamma-1} \nabla \left( s \rho^{\gamma-1} \right) + \lambda(s) \nabla s = 0, $$
gives
		$$ \rho \nabla \phi + \nabla p = 0. $$
The lowest-energy state is thus one in which the gas is in mechanical equilibrium. Moreover, according to Eq.~(\ref{s variation}), in this equilibrium $p$ is a function of $s$ alone, so that the pressure varies with height in the same way at every location on Earth. In other words, as noted by Lorenz, the minimum total energy which can result from adiabatic rearrangement of air occurs when the pressure is everywhere equal to its horizontal average. 
	
\bibliographystyle{jpp}

\bibliography{per}

\end{document}